\pdfoutput=1
\RequirePackage{ifpdf}
\documentclass{PoS_oldRef}

\title{Complex organic molecules in protostellar environments in the SKA era}

\ShortTitle{Complex organic molecules in protostellar environments in the SKA era}

\author{
C. Codella$^1$,\speaker{L. Podio}$^1$, F. Fontani$^1$, I. Jim\'enez-Serra$^2$, P. Caselli$^3$, C. Ceccarelli$^4$,  
M.E. Palumbo$^5$, A. L\'opez-Sepulcre$^{6,4}$, M.T. Beltr\'an$^1$, B. Lefloch$^4$, J.R. Brucato$^1$,
S. Viti$^7$, L. Testi$^{2,1}$ 
\\
$^1$ INAF, Osservatorio Astrofisico di Arcetri, Largo E. Fermi 5, 50125 Firenze (Italy); 
$^2$ ESO, Karl Schwarzschild srt. 2, 85748 Garching (Germany); 
$^3$ MPE, Giessenbachstr.1, 85748 Garching (Germany); 
$^4$ IPAG, UMR 5274, UJF-Grenoble 1/CNRS-INSU, 38041 Grenoble (France); 
$^5$ INAF, Osservatorio Astrofisico di Catania, via S. Sofia 78, 95123 Catania (Italy);
$^6$ Department of Physics, The University of Tokyo, Bunkyo-ku, Tokyo 113-0033, Japan;
$^7$ UCL, Gower Street, WC1E 6B London (UK)
\\
E-mail: \email{codella@arcetri.astro.it}
}

\abstract{Molecular complexity builds up at each step of the Sun-like star formation process, 
starting from simple molecules and ending up in large polyatomic species. Complex organic 
molecules (COMs; such as methyl formate, HCOOCH$_3$, dymethyl ether, CH$_3$OCH$_3$, formamide, NH$_2$CHO, or 
glycoaldehyde, HCOCH$_2$OH) are formed in all the components of the star formation recipe 
(e.g. pre-stellar cores, hot-corinos, circumstellar disks, shocks induced by fast jets), 
due to ice grain mantle sublimation or sputtering as well as gas-phase reactions. 
Understanding in great detail the involved processes is likely the only way to predict 
the ultimate molecular complexity reached in the ISM, as the detection of large molecules 
is increasingly more difficult with the increase of the number of atoms constituting them.

Thanks to the recent spectacular progress of astronomical observations, due to the 
Herschel (sub-mm and IR), IRAM and SMA (mm and sub-mm), and NRAO (cm) telescopes, an enormous 
activity is being developed in the field of Astrochemistry, extending from astronomical 
observatories to chemical laboratories.
We are involved in several observational projects providing  
unbiased spectral surveys (in the 80-300 and 500-2000 GHz ranges) with unprecedented sensitivity of 
templates of dense cores and protostars. Forests of COM lines have been detected. 
In this chapter we will focus on the chemistry of both cold prestellar cores and hot
shocked regions, (i) reviewing results and open questions provided
by mm--FIR observations, and (ii) showing the need of carrying on the observations of
COMs at lower frequencies, where SKA will operate. We will also emphasize the importance of analysing 
the spectra by the light of 
the experimental studies performed by our team, who is investigating the chemical effects induced 
by ionising radiation bombarding astrophysically relevant ices.}

\FullConference{
Advancing Astrophysics with the Square Kilometre Array\\
June 8-13, 2014\\
Giardini Naxos, Italy}

\newcommand{\skipthis}[1]{}

\begin{document}
\makeatletter
\setbox\@firstaubox\hbox{\small Claudio Codella}
\makeatother

\section{The molecular complexity in a Sun-like forming system}

The formation of Sun-like stars and the chemical
complexity of the molecular gas involved in the process
are sketched in Fig. 1, following
Caselli \& Ceccarelli (2012), and here summarised:

\begin{enumerate}

\item
matter slowly accumulates toward the
center of a molecular
cloud. The central density increases while the temperature decreases, forming
the so-called prestellar cores.
Atoms and molecules in the gas phase freeze-out onto the cold
surfaces of the dust grains, forming the grain mantles.
Hydrogenation of atoms and molecules takes place, forming molecules such
as water (H$_2$O), formaldehyde (H$_2$CO) and methanol (CH$_3$OH).
In these regions the formation of new molecules in icy mantles is also caused by
the effects of UV photons and low-energy cosmic rays.
Indeed, the recent detection of CH$_3$CHO, CH$_3$OCH$_3$, CH$_3$OHO, and CH$_2$CO in the 
prestellar core L1689B (Bacmann et al. 2012) shows that COM synthesis has already
started at the prestellar stage.

\item
The collapse starts, the gravitational energy
is converted into radiation and the envelope around the central object
warms up.
The molecules frozen on the mantles acquire mobility and
form new, more complex species. When the temperature reaches
about 100 K mantle sublimates, and we have
the so called hot corinos ($\le$ 0.01 pc) phase.
Molecules in the mantles are injected in the gas, where they
react and form new, more complex, molecules.
The abundance of COMs (such as methyl formate, HCOOCH$_3$, or
dymethyl ether, CH$_3$OCH$_3$) dramatically increases.
A classical example is provided by IRAS16293-2422 
(e.g. Cazaux et al. 2003; Ceccarelli
et al. 2007; Bottinelli et al. 2007), where recently also glycoaldehyde
(HCOCH$_2$OH), crucial molecules for the formation of metabolic molecules,
has been detected (J\"orgensen et al. 2012).

\item
Simultaneously to the collapse, a newborn protostar generates a
fast and well collimated jet, possibly
surrounded by a wider angle wind. In turn, the ejected material drives
shocks travelling through the surrounding high-density medium.
Shocks heat the gas up to thousands of K and trigger several processes such as
endothermic chemical reactions and ice grain mantle sublimation
or sputtering. Several molecular species undergo significant
enhancements in their abundances.
The prototypical chemical rich shock is L1157-B1.
Towards this source, not only relatively simple complex molecules, like methanol,
have been detected (Bachiller et al. 2001), but also molecules considered hot corinos tracers,
like methyl cyanide (CH$_3$CN), ethanol (C$_2$H$_5$OH),
formic acid (HCOOH), and HCOOCH$_3$ (Arce et al. 2008).
The emission of these species is concentrated in a small (around 1000 AU) region
associated with the violent shocks at the head of the outflowing
material (Codella et al. 2009). The presence of COMs in molecular outflows
strongly suggests that these species were part of the sputtered icy mantles
as the time elapsed since the shock is too short for any gas-phase route to build up COMs.

\item
The envelope dissipates with time and eventually only a circumstellar disk remains,
also called protoplanetary disk. In the hot regions, close to the central forming
star,
new complex molecules are synthesized by reactions between the species
formed in the protostellar phase. In the cold regions of the disk,
where the vast majority of matter resides, the molecules formed
in the protostellar phase freeze-out onto the grain mantles again.
Dust grains then coagulate into larger planetesimals, the bricks of
future planets, comets, and asteroids.

\end{enumerate}

\begin{figure}
\includegraphics[angle=0,width=1.0\textwidth]{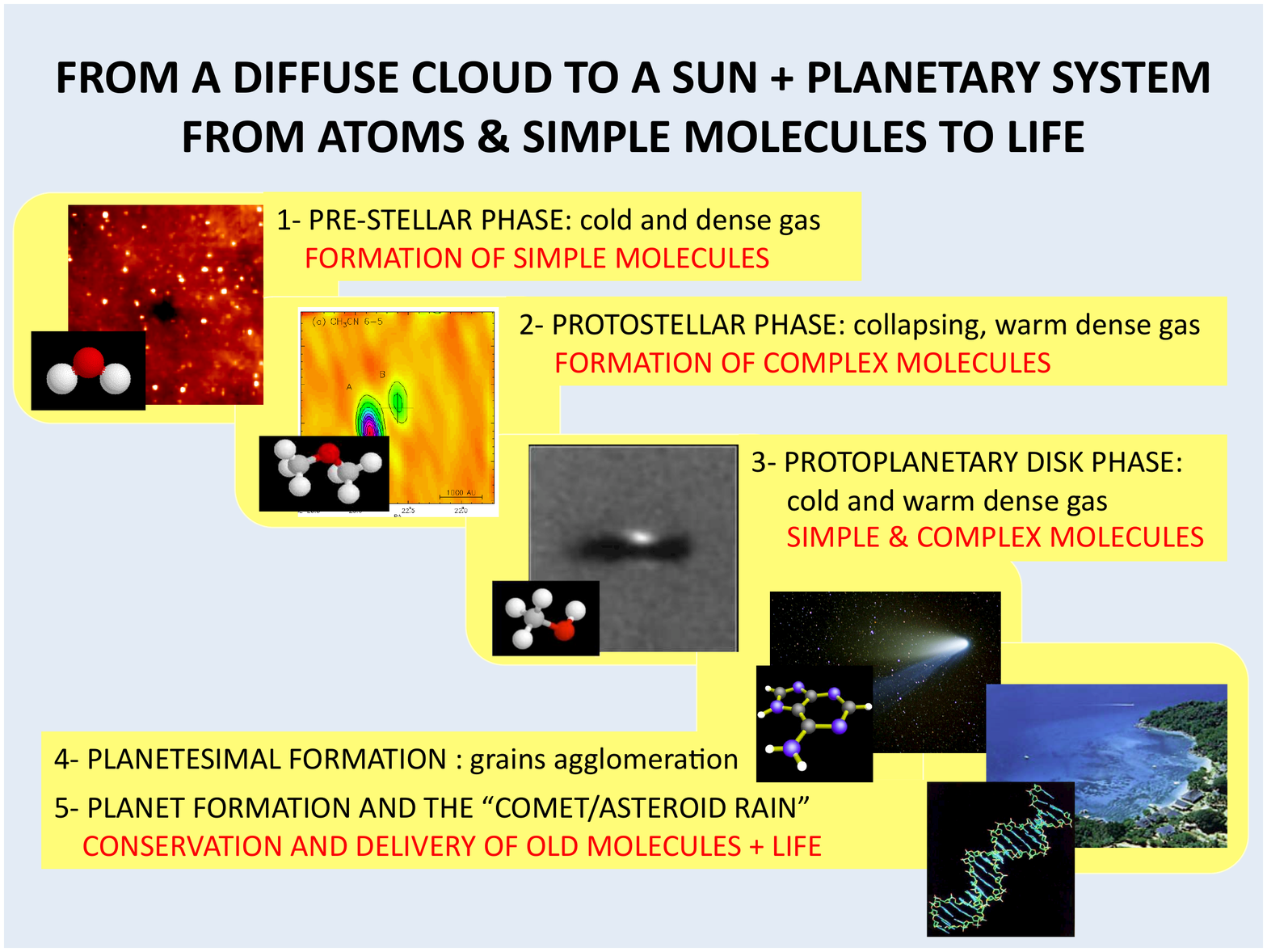}
\caption{Star formation and chemical complexity
(from Caselli \& Ceccarelli 2012). The formation of a star and a planetary
system, like the Solar System, passes through different phases, marked in the sketch.}
\label{life}
\end{figure}

In this chapter, we focus on the earliest stages reviewing the importance of COMs
in the prestellar cores as well as in the hot shocked regions produced by protostellar jets.
Planetary disks will be discussed in a separate chapter (Testi et al. 2014).  

\section{The lessons provided by mm--FIR unbiased spectral surveys}

In the last years several large programs started collecting 
unbiased spectral surveys with high frequency resolutions and unprecedented sensitivities
of different targets considered among the typical laboratories of
different stages of the low-mass star forming process (as e.g. prestellar cores,
hot-corinos, protostellar shocks, more evolved Class I objects).
One of the main goals is indeed the detection of complex and rare molecular species
in the interstellar space through emission due to their ro-vibrational
transitions.
In particular, these efforts have been recently carried out
in the 80--300 GHz range using the IRAM 30-m ground based observatory 
(ASAI: Astrochemical Surveys At IRAM; http://www.oan.es/asai)
and between 500 GHz and 2000 GHz using the ESO Herschel Space Observatory 
(CHESS: Chemical HErschel Surveys of Star forming regions; http://www-laog.obs.ujf-grenoble.fr/heberges/chess/). 
The collected spectra are very rich in COMs, reflecting the chemical complexity of all
the components involved in the low-mass star formation, and, in the case of 
hot-corinos around protostars and jet-driven shocks amazing forest 
of lines are detected (e.g. Ceccarelli et al. 2010).
The analysis of the large number of emission lines due to COMs is still in progress, but the preliminary
results are very encouraging: for instance formamide (NH$_2$CHO),  
the simplest possible amide and a central molecule in the 
synthesis of metabolic and genetic molecules, has been revealed towards
IRAS16293-2422, NGC1333-IRAS4A, SVS13A, OMC-2 FIR4, CephE, and the L1157-B1 shock 
(Kahane et al. 2013; Mendoza et al. 2014; L\'opez-Sepulcre et al. 2014, see Fig. 2). 

\begin{figure}
\centering
\includegraphics[width=0.8\textwidth]{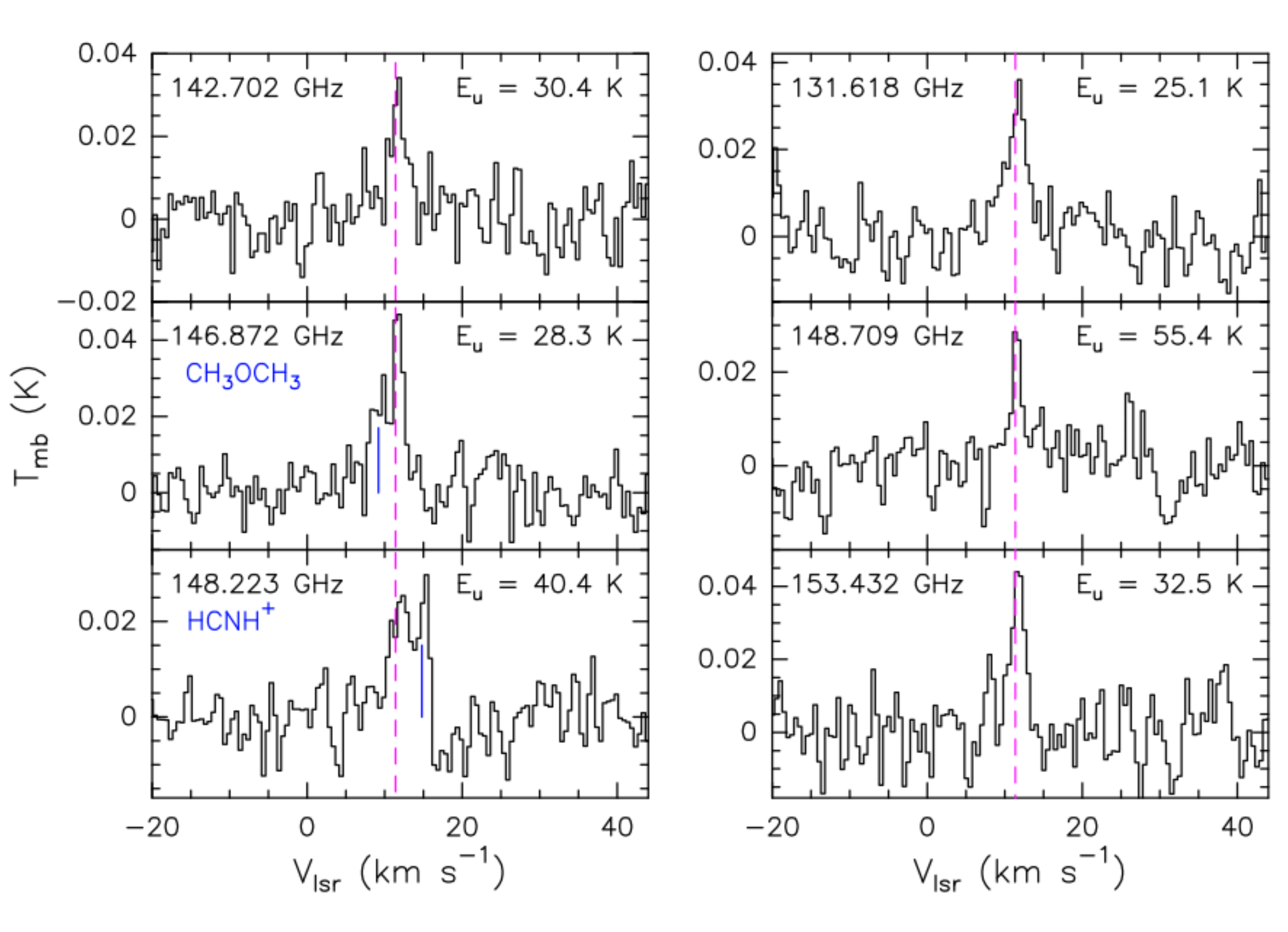}
\caption{NH$_2$CHO lines detected at 2 mm in OMC-2~FIR4 
with the IRAM 30-m antenna in the ASAI framework
(from L\'opez-Sepulcre et al. 2014). 
The magenta dashed lines mark the systemic velocity of OMC-2~FIR4. 
The blue solid lines mark the position of blended lines.}
\label{formamide}
\end{figure}

A huge effort is carried out also using interferometers to provide high-spatial
resolution images. The IRAM Plateau de Bure interferometer (PdBI) large program CALYPSO
(Continuum and Lines from Young ProtoStellar Objects; http://irfu.cea.fr/Projects/Calypso) is providing the 
first sub-arcsecond statistical study, in the 80--300 GHz window, of the inner 
environments of the low-luminosity Class 0 sources.
Also in this case spectacular forests of lines are observed,
showing an amazing large number of lines at high
excitation due to COMs such as ethylene glycol ($aGa^{'}$--(CH$_2$OH)$_2$) and
NH$_2$CHO. As an example, Fig. 3 shows the NGC1333--IRAS2A case (Maury et al. 2014): COM emission inside
the protostellar envelope has been imaged, finding that
it originates from a region of radius 40--100 AU, centered on the protostar.
This spatial scale is consistent 
with the size of the inner envelope where $T_{\rm dust}$ $\geq$ 100 K is expected, supporting
the hot-corino origin.

In conclusion, the synergy between spectral surveys using single-dishes and  
interferometric observations is of paramount importance to detect and analyse the COM emission.
On the one hand, the unbiased spectral surveys provide the largest possible
frequency range, and thus the most complete census of COMs; a multiline approach
is also needed to safely identify the emission spectrum of complex species. 
On the other hand, interferometric images provide the size and morphology of the emitting
regions, overcome the filling factor limitation of single-dish spectra, and allow one
to propely derive physical conditions such as density, temperature, and abundance.
In addition, the linewidths of COMs are typically reduced in interferometric observations, 
which allows a better identification of the molecular lines since it reduces line blending. 

\begin{figure}
\centering
\includegraphics[width=0.9\textwidth]{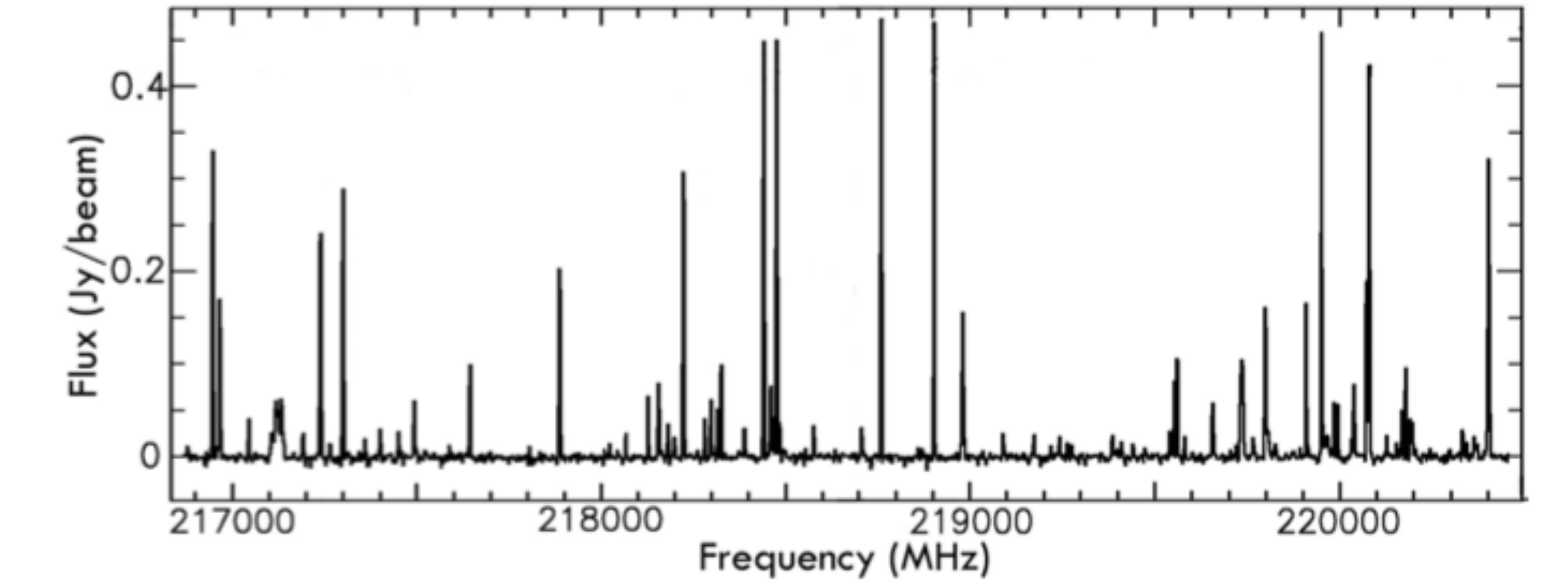}
\caption{Continuum-subtracted PdBI spectrum around 218.5 GHz, at the position of the IRAS2A protostar 
(from Maury et al. 2014).
Several COMs are detected, as e.g. CH$_3$OCH$_3$, CH$_3$OCHO, $aGa^{'}$--(CH$_2$OH)$_2$, and NH$_2$CHO.}
\label{calypso}
\end{figure}

\section{Cold cases: prestellar cores and very early protostars}

\subsection{Detectability of aminoacids in Solar-system precursors}

The increasing performance of millimeter instrumentation in the past years has opened up 
the possibility to carry out high-sensitivity molecular line surveys even toward the earliest 
(and coldest) stages of low-mass star formation. These initial conditions are represented 
by cold dark cores and in particular, by pre-stellar cores, i.e. dense and cold 
condensations on the verge of gravitational collapse (central H$_2$ densities of some 10$^4$ cm$^{-3}$ 
and temperatures $\leq$ 10 K; Caselli et al. 1999; Crapsi et al. 2007). These observations have shown 
that COMs (such as propylene - CH$_2$CHCH$_3$ - or acetaldehyde - CH$_3$CHO) are also present 
in the cold gas of these cores, unexpectedly revealing a high chemical complexity in these objects 
(see Marcelino et al. 2007 and Bacmann et al. 2012).

Among these complex organics, amino acids are of high interest in Astrochemistry and Astrobiology 
because of their role in the synthesis of proteins in living organisms. It is currently believed 
that their formation may have occurred in the interstellar medium (ISM) since amino acids, including 
glycine and alanine, have been found in meteorites (Pizzarello et al. 1991; Ehrenfreund et al. 2001; 
Glavin et al. 2006) and comets (as in Wild 2; Elsila et al 2009). This idea is supported by laboratory 
experiments which have shown that amino acids precursor are largely produced in UV-photon 
and ion-irradiated interstellar ice analogs (Mun\~oz-Caro et al. 2002; Bernstein et al. 2002; Holtom et al. 2005). 
The detection of amino acids in the ISM is however challenging since their partition function is large 
and therefore, their emission is spread among many transitions. The detection of amino acids in the ISM remains 
to be reported.

Recently, it has been proposed that pre-stellar cores may be well suited for the detection of 
amino acids in the ISM (Jim\'enez-Serra et al. 2014). The gas temperatures in pre-stellar cores 
are $\leq$ 10 K, which yields a low level of line confusion since the number of molecular 
transitions excited at these temperatures is small. The linewidths of the molecular line emission 
in these cores are $\leq$ 0.5 $\,$km$\,$s$^{-1}$, which allows accurate identifications of the 
observed molecular lines because they suffer less from line blending. In addition, water vapour 
has recently been found toward the central few thousand AU of one pre-stellar core, L1544, which 
indicates that a small fraction of the ices ($\sim$ 0.5\% of the total water abundance in the mantles) 
has been released into the gas phase (Caselli et al. 2012). These authors have proposed that water 
vapour in L1544 is produced by the partial photo-desorption of ices by secondary, cosmic ray induced 
UV-photons. Since COMs form in the outer layers of the mantles (Cuppen et al. 2009; Taquet et al. 2012), 
these molecules are also expected to be photo-desorbed together with water in pre-stellar cores, making 
these objects excellent candidates to test the detectability of amino acids in Solar-system precursors.

Simple radiative transfer calculations of glycine (NH$_2$CH$_2$COOH, i.e. the simplest amino acid) 
towards the L1544 pre-stellar core show that several glycine lines could reach detectable levels 
(peak line intensities $\geq$ 10 mK) in the frequency range between 60 and 80 GHz 
(Jim\'enez Serra et al. 2014). We have extrapolated these results to the frequency range covered by 
SKA1-MID Bands 4 and 5, where we evaluate whether glycine could be detected with this instrument 
(see Fig. 4). These calculations assume a solid abundance of glycine in ices of a few 
10$^{-4}$ with respect to water, similar to those synthesized in laboratory experiments of 
UV photon- and ion-irradiated interstellar ice analogs (Mun\~oz-Caro et al. 2002; Bernstein et al. 2002; 
Holtom et al. 2005). This solid abundance translates into a maximum gas-phase abundance of glycine of 
$\sim$ 8$\times$10$^{-11}$ after ice photo-desorption.

In Fig. 4, we report the predicted intensities of glycine for the L1544 core, 
assuming that the gas-phase abundance of glycine (of $\sim$ 8$\times$10$^{-11}$) remains 
constant across the core. The velocity resolution used in the simulations is $\sim$ 0.2 $\,$km$\,$s$^{-1}$ 
and the expected linewidth of the glycine emission is $\sim$ 0.3$\,$km$\,$s$^{-1}$. 
The assumed source size is $\sim$ 12". 
Figure 4 shows that a few glycine lines could reach detectable levels with S/N ratios $\geq$ 3 
in a total of $\sim$ 1000 hrs of integration time in SKA1-MID Band 5. 
For this integration time, a rms noise level of 50 microJy is indeed expected 
at a velocity resolution of 0.2 km$\,$s$^{-1}$ within a 12"-beam 
(see the Level 0 Science Requirement SKA1-SCI-19). 
However, we stress that if the frequency coverage of SKA 
were extended to higher frequencies (e.g. up to 25 GHz, which would also allow 
to observe the NH$_3$(1,1) inversion lines at $\sim$ 23 GHz; see Fig. 4), 
this detection experiment could be performed in 10 times less time thanks to the presence of 3 
times brighter glycine lines around $\sim$ 20 GHz (intensities $\geq$ 0.5 mJy beam$^{-1}$). In any case, 
a gas-phase abundance as low as $\sim$ 8$\times$10$^{-11}$ would be the lowest abundance of 
glycine that could be detected with this experiment in SKA1-MID Band 5.

It is important to stress that Bands 1 and 2 of ALMA are currently unfunded, 
and therefore SKA may represent in the future the only instrument 
that could perform this kind of detection experiment for amino acids. 
We also note that although glycine shows a collection of transitions at millimetre wavelengths, 
observations at centimetre wavelengths with SKA1-MID Band 5 (8.8 to 13.8 GHz) are essential 
to provide reliable identifications of these lines due to the increasing frequency span 
between transitions at longer wavelengths (preventing line blending and line confusion). 
In addition, other complex organics such as the precursors of the glycine chemistry, e.g. 
amino acetonitrile or methyl formate, are expected to be factors of $\sim$100 more abundant 
than glycine (Belloche et al. 2008, 2013), and therefore they would be easier to detect 
with SKA1-MID Bands 4 and 5.

The detection of glycine, and of its precursors, in these Solar-system precursors will represent 
a major milestone in Astrochemistry and Astrobiology, providing a unique opportunity to link 
the pre-biotic chemistry in the ISM to their subsequent delivery onto planetary systems.

\begin{figure}
\centering
\includegraphics[width=0.8\textwidth]{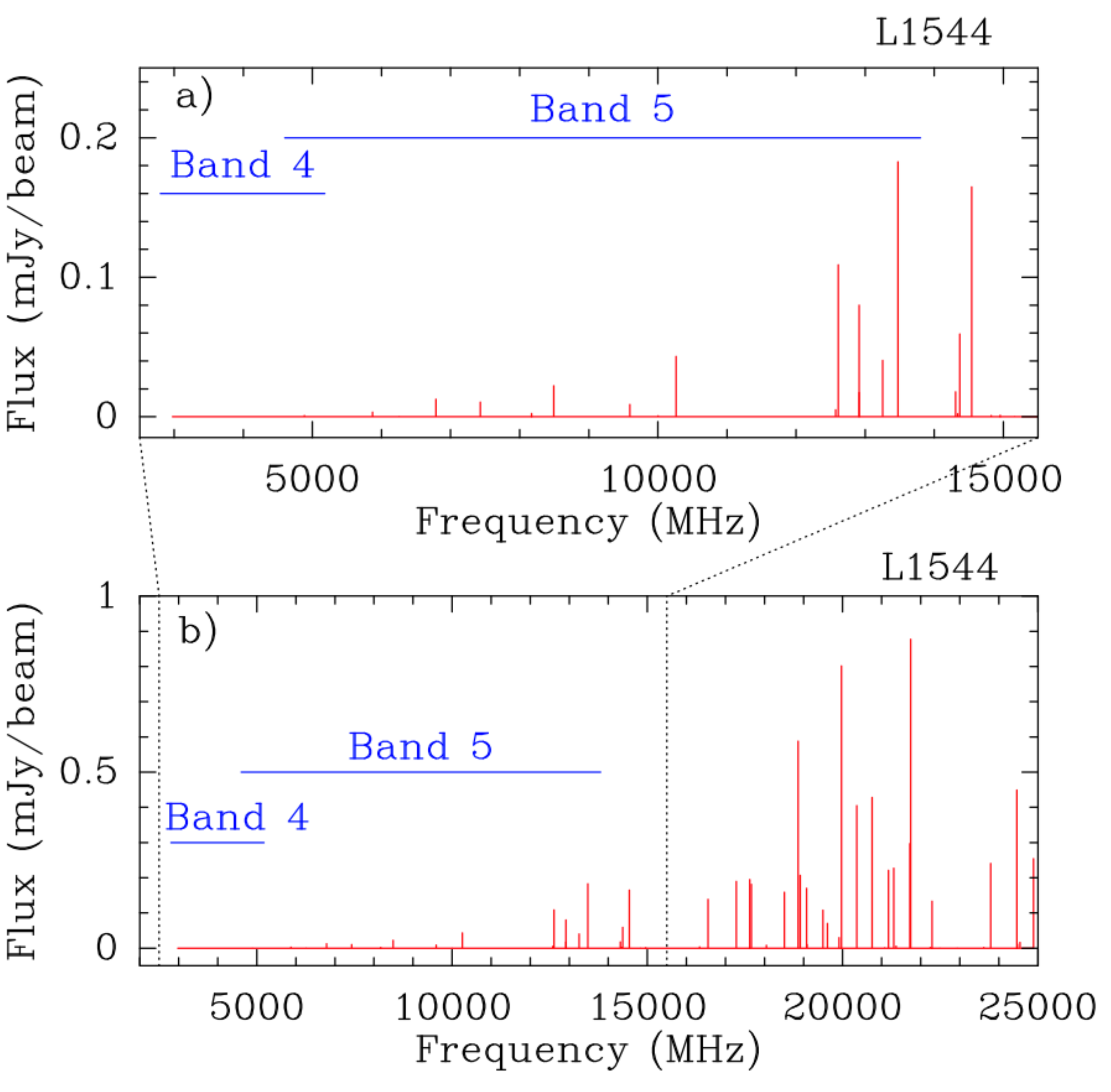}
\caption{Upper panel: Simulations of the spectrum of glycine (conformer I) obtained for the 
frequency range between 2.5 GHz and 15.5 GHz, considering the physical structure 
of the L1544 pre-stellar core (Caselli et al. 2012), a solid glycine abundance of a few 10$^{-4}$ 
with respect to water in ices, and LTE conditions. Horizontal blue lines indicate 
the frequency coverage of SKA1-MID Bands 4 and 5. Lower panel: LTE spectrum of 
glycine predicted for frequencies between 2.5$\,$GHz and 25$\,$GHz. The glycine 
lines around 20 GHz are factors of $\sim$3 brighter than those covered by SKA1-MID Band 5, 
which would allow the detection of this amino acid in 10 times less observing time.}
\label{glycine}
\end{figure}

\subsection{Deuterated COMs}

There are no doubts that one of the most important chemical processes
occuring in the cold (T $\leq 20$ K) and dense ($\geq 10^4$ cm$^{-3}$)
pre--stellar cores is the freeze-out of atoms and molecules on
the surface of dust grains, where the depletion of both C-bearing
and N-bearing molecules is found to be high in low- and high-mass
pre--stellar cores (e.g.~Caselli et al. 1999, Friesen et al. 2010,
Fontani et al. 2012). Frozen on grain mantles, atoms and molecules
can undergo hydrogenation, due to the high mobility of the light H
atoms. In particular, from hydrogenation of CO (the most abundant neutral
molecule after H$_2$) one forms sequentially HCO,
H$_2$CO and CH$_3$OH, which is thus the species that is formed last.
Moreover, the low temperature favours the deuteration
of the species mentioned, because of the endothermicity of
the chemical reactions which replace D atoms
into their hydrogenated counterparts. Therefore, as
time proceeds, the formation of methanol and its deuterated
forms (CH$_2$DOH, CH$_3$OD, CHD$_2$OH, etc.)
is boosted, until the energy released by the nascent
protostellar object in the form of radiation increases
the temperature of its environment, causing the evaporation
of the grain mantles and the release of these molecules
into the gas. As the temperature increases, the deuterated
species are expected to get gradually destroyed due to
the higher efficiency of the backward endothermic reactions
(Caselli \& Ceccarelli 2012).
Therefore, high deuterated fractions of
methanol, i.e. the ratio between the column density of the
species containing D and that of CH$_3$OH, are expected
to be powerful tracers of the short-living evolutionary stage
in between the pre-stellar and the protostellar
phase, at which the evaporation/sputtering of the grain
mantles is efficient, and the warm gas-phase reactions
have no time to change significantly the chemical composition
of the gas.

Of course, testing these predictions is challenging
because the deuterated forms of methanol are supposed to
originate faint lines, which, at (sub-)millimeter wavelengths,
can be easily overwhelmed by nearby stronger emission lines
of lighter and more abundant molecules.
Several transitions of CH$_2$DOH can be observed in the range 1--15~GHz
as can be seen in Fig. 5.
In this spectral window, the contamination of lines of more abundant molecules
is expected to be negligible.
The combination of high angular resolution
and high sensitivity offered by SKA (see Sect. 7) will be eminently
suitable to map these lines, which are expected to arise from
very compact regions (sizes $\leq 3$", e.g. Peng et al. 2012).
As an example, we can estimate the time required
to detected some of the lines shown in the synthetic spectrum in
Fig.~5. The spectrum includes all transitions of CH$_2$DOH in the
spectroscopic band $\sim 1-15$~GHz, as modeled by GILDAS--Weeds package 
(Maret et al. 2011) assuming the following parameters: $T_{\rm kin} = 20$ K,
$N$(CH$_2$DOH) = $5 \times 10^{15}$ cm$^{-2}$, source size = 1",
line FWHM = 1 km s$^{-1}$. The column density assumed is a
mean source-averaged value measured towards 
protostellar cores (e.g. Parise et al. 2006; Fontani et al.~2014b).
Using the SKA1 Imaging Science Performance document 
(Braun R., 2014-06-02 version), at a rest frequency  
of 5~GHz, assuming $T_{\rm sys} = 20$ K, 190 dishes of 15~m, 2 polarisations,
a velocity resolution of $\sim 1$ km s$^{-1}$, the 1$\sigma$~rms
noise in the spectrum after 50~hrs of integration on source is 0.077 mJy.
Figure 5 tells us that several lines have intensity peak well above
3$\sigma$. Therefore, this will allow us not only to just detect the
presence of the molecule, but also to derive estimates of some important
physical parameters (e.g. rotation temperature and column density
from the rotation diagrams).

Moreover, other deuterated species of
complex molecules (CH$_2$DCCH, HDNCH$_2$CN, etc.) possess
lines in the same frequency range. If detected, these species will
put other relevant constraints on chemical models, in an effort to
reach a general knowledge of the deuteration mechanism(s) during
the earliest phases of the star formation process.

\begin{figure}
\centering
\includegraphics[angle=0,width=0.9\textwidth]{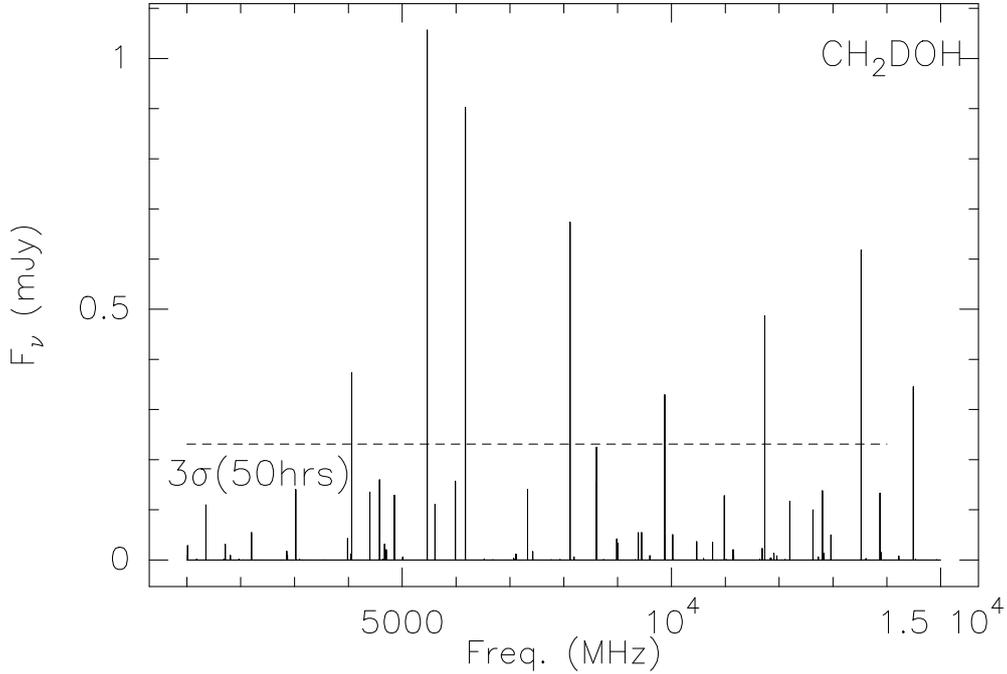}
\caption{Synthetic spectrum of CH$_2$DOH modeled for SKA1-MID with GILDAS--Weeds package 
(Maret et al. 2011) in the range $\sim$ 1--15 GHz, assuming 
LTE conditions, $T_{\rm kin} = 20$ K,
$N$(CH$_2$DOH) = $5 \times 10^{15}$ cm$^{-2}$, source size = 1",
and line FWHM = 1 km s$^{-1}$. The dashed line represents the
expected 3$\sigma$ level in the spectrum that can be achieved after
50 hours of integration on source (see text for details).}
\label{deute}
\end{figure}

\section{Hot protostellar shocks versus hot-corinos}

The L1157 region at 250 pc hosts a
Class 0 protostar (L1157-mm) driving a spectacular chemically rich
bipolar outflow (Bachiller et al. 2001). 
The southern lobe is associated with
two cavities seen in the IR H$_2$ and CO lines
(e.g. Neufeld et al. 2009; Gueth et al. 1996),
likely created by episodic events in a precessing jet.
Located at the apex of the more recent cavity, the bright bow
shock called B1 has a kinematical age of 2000 years.
The bow shocks, when mapped with the IRAM PdB and
VLA interferometers, reveal a clumpy structure, with clumps
located at the wall of the cavity (e.g. Tafalla \&
Bachiller 1995; Codella et al. 2009), and well traced
by typical products of both grain mantle sputtering (such as 
NH$_3$, CH$_3$OH, and H$_2$CO) and refractory core disruption (SiO). 
Thus, the young L1157-B1 shock offers  
an exceptional opportunity to investigate in details the chemical
composition of the grain ice mantles as well as how the gas phase 
is chemically enriched after the shock occurrence.

Thanks to PdBI observations (see Fig. 6), three COMs have been so far imaged in L1157-B1:
CH$_3$OH, acetaldehyde (CH$_3$CHO), and CH$_3$CN (Benedettini et al. 2007, 2013;
Codella et al. 2009, 2014).
The CH$_3$CN image shows a clumpy structure superimposed to the classical B1 arch-like shape, 
displaying a unique continuous structure tracing the propagation of a large bow shock.
In principle, CH$_3$CN could form on the surface of grains and then 
injected in the gas phase due to shock sputtering.
Alternatively, CH$_3$CN may belong to the so-called 
second generation species, i.e. the molecules formed in the warm 
gas-phase chemically enriched after the mantle release 
(see e.g. Bottinelli et al. 2007; Bisschop et al. 2008).
Both scenarios need an increase of temperature,
as in the L1157-B1 case (up to about 10$^3$ K, Busquet et al. 2013).
The CH$_3$CN-to-CH$_3$OH ratio 
is $\simeq$ 10$^{-4}$--10$^{-3}$, i.e. smaller 
with respect to those found in hot-corinos (10$^{-2}$--10$^{-1}$).
This difference suggests that a significant 
amount of CH$_3$CN may form in the gas phase and that in 
L1157-B1 the CH$_3$CN abundance has 
not yet reached its maximum, given the young age of the shock.

Recently, also CH$_3$CHO has been mapped at PdBI (Codella et al. 2014),
and also in this case there is a good agreement with the CH$_3$OH
(and CH$_3$CN) spatial distribution, confirming COMs are emitting
where mantles have been recently sputtered.
This is further indicated by the excellent spatial correpondance
with HDCO emission (Fontani et al. 2014a) tracing deuterated formaldehyde
formed on grains and then injected in the shocked gas.
For acetaldehyde, the abundance ratio with respect to methanol is
quite high ($\sim$ 10$^{-2}$) and in any case similar to that
observed towards a typical hot-corino such as IRAS16293-2422 (Bisschop et al. 2008). 
These findings support for CH$_3$CHO either a direct formation on grain mantles or
a quick ($\le$ 2000 yr) formation in a chemically enriched gas phase.  

These first results show the importance of studying shocked regions as
laboratories where (i) to investigate the COMs formation routes 
and (ii) to verify whether the study of hot-corinos chemistry can
be affected by COM emission
arising from shocked envelope material at the base of the inner (unresolved) jet. 
Interferometric observations are instructive to minimise  
beam dilution effects. In addition, the multiline approach 
is needed to carefully sample the excitation conditions and correctly 
derive the abundances of COMs. In particular, the 1--15 GHz frequency range
allows one to well sample low-excitation (E$_u$ $\le$ 20 K) and bright
(S$\mu^2$ $\geq$ 1 D$^2$) emissions due to COMs (see Sect. 7). 

\begin{figure}
\centering
\includegraphics[angle=0,width=0.9\textwidth]{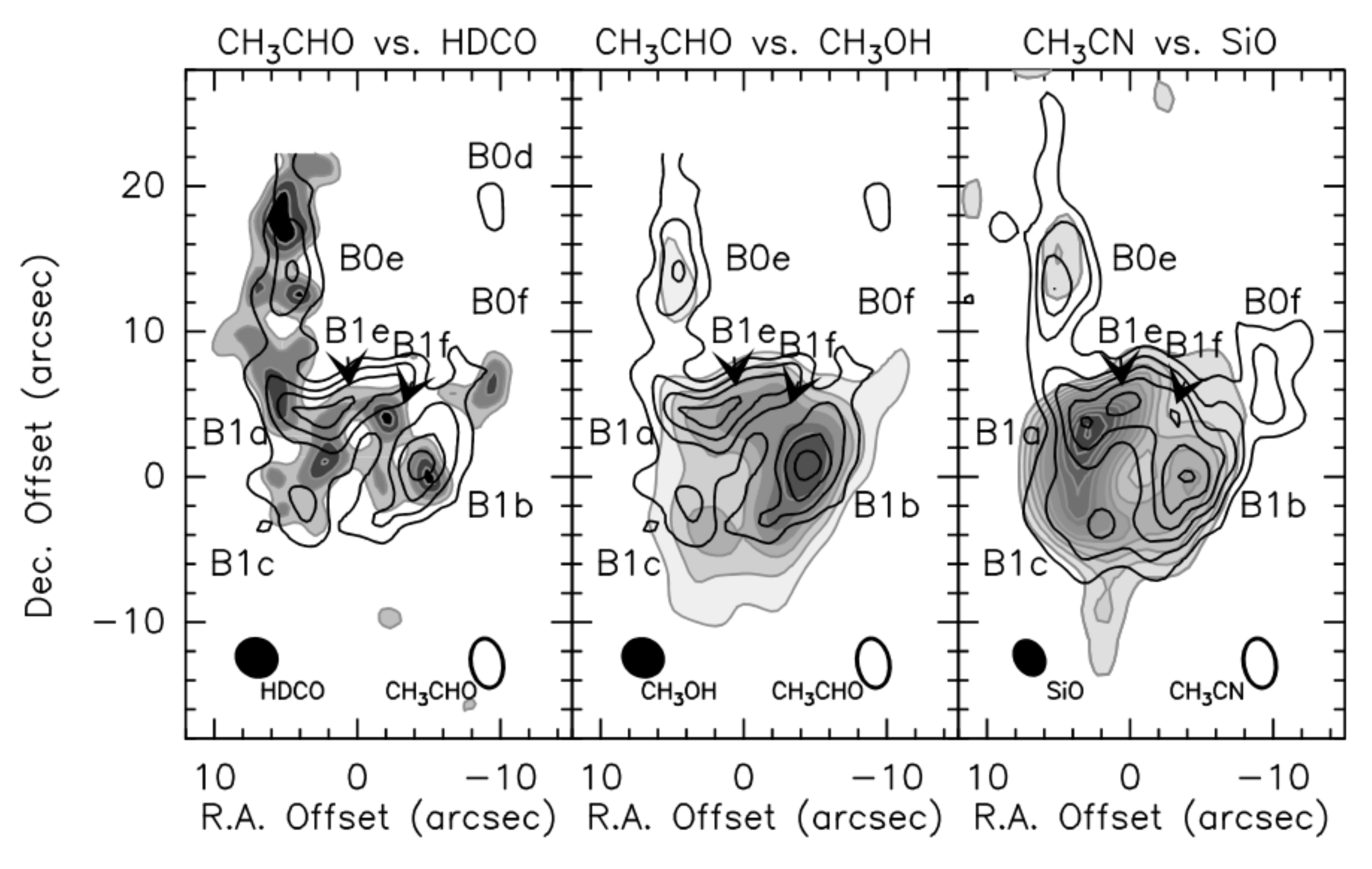}
\caption{Chemical differentiation in L1157-B1: the maps are centred at
$\Delta\alpha$ = +21$^{o}$ and $\Delta\delta$ = --64$^{o}$
from the driving protostar L1157-mm. 
Overlay of the integrated intensity IRAM PdBI map at 2 mm of the 
CH$_3$CHO(7$_{0,7}$--6$_{0,6}$)AE (contours, from Codella et al. 2014)
on the HDCO(2$_{1,1}$--1$_{0,1}$)  
(left panel, from Fontani et al. 2014a) and CH$_3$OH(3$_{K}$--2$_{K}$) 
emission (middle panel; from Benedettini et al. 2013).
Righ panel compares the CH$_3$CN(8$_{K}$--7$_{K}$) and SiO(2--1)
spatial distributions (Gueth et al. 1998, Codella et al. 2009).
Labels are for the different L1157-B1 molecular clumps well
imaged in the CH$_3$CN map. 
The synthesised beams are shown in the bottom part of the panels.}
\label{L1157}
\end{figure}

\section{The SKA case for high-mass star formation}

While the detection and study of COMs in low-mass star forming regions has a clear impact on our understaning of the origin of life, 
we should underline that COMs in massive star forming regions have recently proved to be ideal 
tracers of the physical conditions of the very central compact core where the massive star forms. In this subsection we therefore
briefly make a case for why SKA is the ideal instrument to study COMS in high-mass star forming regions.
The earliest stages of the massive star formation are characterised by the occurrence of the so
called hot cores, i.e. hot ($\ge$ 100 K), small ($\le$ 0.1 pc), and massive (up to thousands of solar masses) clumps warmed by the
stellar radiation. In fact, the largest body of detections of COMs comes from observations of hot cores.
In the standard paradigm, at densities as high as those of hot cores, gases freeze onto the surface of dust grains in the cold
quiescent phase of the molecular cloud. This material is processed to greater
molecular complexity via a number of possible surface reactions, perhaps catalyzed
by UV, cosmic rays or thermal effects, and is subsequently released back into
the gas phase as a result of thermal sublimation due to the heating effect of
the newly-formed star.
The rich chemistry, comprising of species such as ethyl cyanide, glycoaldehyde and its isomers (e.g. Beltr\'an et al. 2009,
Fuente et al. 2014, Calcutt et al. 2014, and references therein), that results from this sublimation, 
thus potentially provide powerful diagnostics of the physical and chemical history of the very central core where the star is born. 
Interestingly, and somewhat
surprisingly, hot cores differ from hot corinos not only for the larger sizes,
but also chemically: when normalized to methanol or formaldehyde,
hot cores have typically one order of magnitude less abundant COMs (such as HCOOCH$_3$ or CH$_3$OCH$_3$)
than hot corinos (e.g. Herbst and van Dishoeck 2009). Such differences
are likely due to various factors, such as  the composition of the sublimated ices, governed by the past prestellar history.
COMs in hot cores are even more of a powerful
star formation diagnostics because they are excellent tracers of the very central (often not spatially resolved)
part of the core, where the protostar is born; their detection is therefore a confirmation of the high density warm cores
and, most importantly, their emission is so compact that they must trace the
most central region of the hot core, close to the YSOs. Clearly
interferometric observations are essential in order to resolve such small scale structures.

Despite tracing such a small extent of gas,
it is found (e.g. Calcutt et al. 2014) that there are significant
variations in abundances and abundance ratios across COMs, making these species also
excellent tracers of chemical and physical differentiation across the different classes of objects.
Of particular interest in the study of COMs are the three isomers of
C$_2$H$_4$O$_2$, all of which have now been detected in star-forming regions.  Acetic acid and glycolaldehyde are
particularly interesting due to their significant prebiotic significance, since the former is linked to the formation of ribose and the
latter is only an amine group (NH$_2$) away from glycine, the simplest amino acid.
However, so far the identification and characerization of COMs in hot cores
have been hampered by the incredibly high number of lines in the sub-millimeter spectra: the latter suffer from strong line confusion due to
(i) the richness of the spectrum: the higher the gas density and temperature, the more crowded the spectra will be; 
(ii) blending due to also the large linewidths, and (iii) uncertainties in the laboratory
rest frequencies as well as in the observations. These issues lead, in some cases, to only tentative detections
and contribute to the scarcity of observed transitions and number of objects.

The SKA frequency range offers a unique opportunity to identify and study a large number of COMs
in high mass star forming regions: the frequency range covered
is clear of contamination because the low energy
of smaller molecules fall at higher frequencies, hence line confusion is minimized.
Interferometric observations are very useful in hot cores 
since they filter out much of the extended envelope and therefore the linewidths of COMs can be 
significantly reduced allowing a better identification of the lines.
Also, COMs are characerized by the fact that their 
level population covers many energy levels because of their large partition functions and hence several 
of their transitions fall in the SKA range from 4 GHz to $\sim$ 25 GHz.
As an example, within the 4 GHz limit, there are tens of glycolaldehyde 
and ethylene glycol transitions covering rotational 
transitions from $J$ as low as 1 (for ethylene glycol) and 4 (for glycolaldehyde) up to $J = 30$. 
Clearly, by removing the problem of line confusing and simultaneously observing a large number 
of transitions for each COM we will be able to accurately derive column 
densities, and hence, with the aid of state of the art chemical and radiative transfer modelling, 
determine the physical conditions of the most compact cores in high-mass star forming regions.
The combination of all these requirements are only met by the SKA.

\section{The need of laboratory experiments}

In dense molecular clouds ($n>10^4$ cm$^{-3}$) gas phase species condense on silicatic 
and carbonaceous grains  giving rise to icy mantles. These mantles are made of molecules which 
directly freeze out from the gas phase (such as CO) and molecules which are formed after grain 
surface reactions (such as H$_2$O). The presence of icy grain mantles is indirectly deduced 
from depletion of gas phase species and is observed in the infrared from absorption features 
attributed to vibrational modes of solid phase molecules superposed to the background stellar spectrum. 
Ices have been observed in star forming regions (both low- and high-mass young stellar objects) 
as well as in quiescent dense clouds. It is widely accepted 
that icy grain mantles are continuously processed by low-energy cosmic rays, electrons, and UV photons.

Ten different molecular species have been firmly identified in interstellar icy grain mantles 
(e.g. Gibb et al. 2004). In particular, water (H$_2$O), carbon monoxide (CO), carbon dioxide (CO$_2$), 
methanol (CH$_3$OH), methane (CH$_4$),  carbonyl sulphide (OCS), formaldehyde (H$_2$CO), 
formic acid (HCOOH), cyanate ion (OCN$^-$) and ammonia (NH$_3$).

Most of the knowledge on the physical and chemical properties of ices is based on the comparison 
between observations and laboratory experiments performed at low temperature (10-80 K). 
Laboratory infrared spectra of icy samples show that the profile of absorption bands depends 
on different parameters such as the other molecules each species is mixed  with and the 
temperature (e.g. Sandford et al. 1988; Ehrenfreund et al. 1999; Palumbo \& Baratta 2000; \"{O}berg et al. 2007).
The properties of icy samples also depend on processing by fast ions, electrons, and UV photons  
(e.g.  Grim \& Greenberg 1987; Palumbo \& Strazzulla 1993; 
Hudson et al. 2001; Strazzulla et al. 2001; Baratta et al. 2002;
Bennet et al. 2007; \"{O}berg et al. 2009; Islam et al. 2014). 
Energetic ions (keV-MeV) passing through molecular solids release energy to the target along 
the ion track. As a consequence molecular bonds are broken and radicals and molecular fragments 
recombine giving rise to molecular species not present in the original sample. Furthermore the 
structure and morphology of the sample is also modified (e.g. Palumbo 2006; Raut et al. 2001). 
In the case of UV photolysis the energy is released to the target material through single 
photo-dissociation or ionization events per incoming photon. Thus after ion bombardment and 
UV photolysis the chemical composition and the structure of the sample is modified. Both more 
volatile and less volatile species are formed and if  C-bearing species are present in the 
original sample a refractory residue is left over after warm-up at room temperature 
(e.g. Moore et al. 1983; Foti et al. 1984; Palumbo et al. 2004).

In the past 35 years several experiments have been performed to study the effects of ion bombardment 
and UV photolysis on the chemical composition of icy samples.
As an example, laboratory experiments have shown that after energetic processing of methanol 
the column density of pristine methanol decreases and new bands appear in the infrared spectra 
indicating the formation of other, also complex, molecules.
Species clearly identified are carbon monoxide (CO), carbon dioxide (CO$_2$), methane (CH$_4$), 
formyl radical (HCO), formaldehyde (H$_2$CO), ethylene glycol (C$_2$H$_4$(OH)$_2$), methyl 
formate (HCOOCH$_3$), and glycolaldehyde (HCOCH$_2$OH) (e.g. Moore et al. 1996; Palumbo et al. 1999,
Hudson \& Moore 2000; Bennett et al. 2007; \"{O}berg et al. 2009; Modica \& Palumbo 2010). 
Figure 7 shows the infrared spectrum in the 
1180-980 cm$^{-1}$ (8.47-10.20 $\mu$m) range of a mixture CO:CH$_3$OH bombarded with 200 keV 
protons at 16 K (Modica \& Palumbo 2010).

\begin{figure}
\centering
\includegraphics[width=0.8\textwidth]{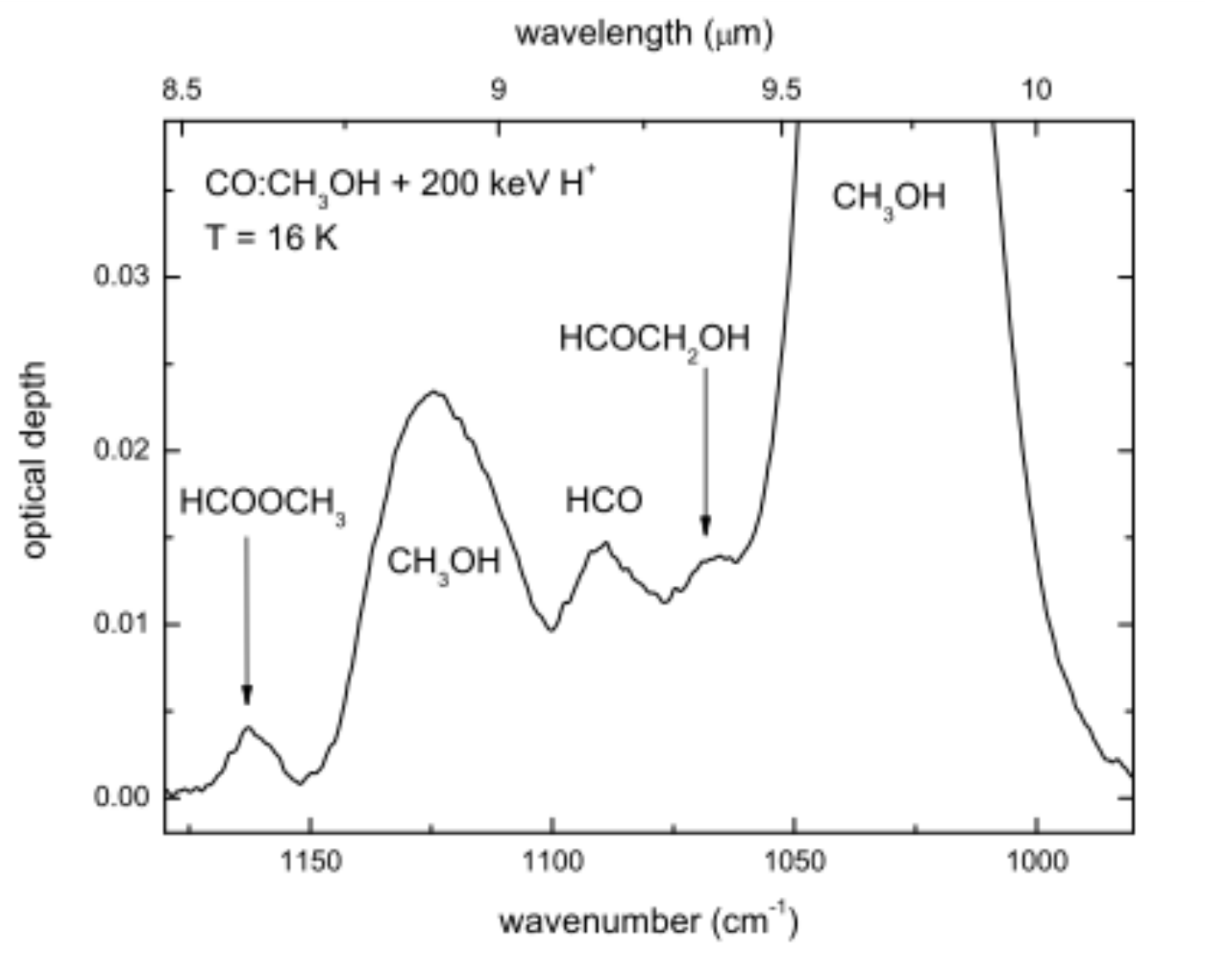}
\caption{Infrared transmission spectrum (from Modica \& Palumbo 2010), 
in optical depth scale, of a mixture CO:CH$_3$OH bombarded 
with 200 keV protons at 16 K. Labels indicate absorption bands due to pristine methanol 
and bands due to species formed after energetic processing.}
\label{lab}
\end{figure}

Recently, Compagnini et al. (2009) and Puglisi et al. (2014) have shown, by infrared and Raman spectroscopy, 
that polyynes (-(C$\equiv$C)$_n$-) and polycumulenes (=(C=C)$_n$=) are formed after ion 
bombardment of  acetylene (C$_2$H$_2$), ethylene (C$_2$H$_4$), ethane (C$_2$H$_6$), and 
benzene (C$_6$H$_6$).
Jones \& Kaiser (2013) have studied the effects of electron irradiation of pure CH$_4$ by reflectron 
time-of-flight mass spectrometry and have found that high-molecular-weight hydrocarbons 
of up to C$_{22}$, among them alkanes, alkenes and alkynes, are formed.
Other experiments (e.g. Kobayashi et al. 1995; 
Mun\~oz-Caro et al. 2002; Bernstein et al. 2002; Holtom et al. 2005) have shown, by high performance 
liquid chromatography, that amino acid precursors are present in the residue formed after 
energetic processing of simple ice mixtures.

Based on these experimental results it has been suggested that icy grain mantles are 
constituted not only by the molecular  species that have been firmly identified but also by 
other, more complex, molecules  which cannot be detected in the solid phase by IR spectroscopy. 
These species are expected to enrich the gas phase composition after desorption of icy 
grain mantles (e.g. Palumbo et al. 2008; Modica \& Palumbo 2010) and could be incorporated in planetesimals and comets.

Finally, these results support the experimental effort (e.g. Allodi et al. 2013) to use more 
sensitive techniques to evidence the formation of complex molecules and/or fragments that 
could be of primary relevance for Astrobiology also to understand which species should be 
searched for, by ground-based or space-born facilities, in protostellar environments, 
protoplanetary disks and in the atmospheres of extrasolar planets and moons.

\section{Why SKA?}

The numerous detections of high-excitation COM lines in the mm-window
call for observations at lower wavelengths, where
heavy species are expected to emit (JPL, http://spec.jpl.nasa.gov, and CDMS,
http://www.astro.uni-koeln.de/cdms).
Even more important, the millimeter frequency bands are often
so full of lines that it is paradoxically difficult to
identify species through a large number of emission lines simply due to confusion.
On the other hand, this low frequency band is relatively clear, given that low energy transitions
of light molecules fall at much higher frequencies.
The completion of the results obtained at mm-wavelengths with spectral surveys in the cm-window
will allow one to have the possibility to have, for different species, 
a large number of transitions, which is needed to reliably
detect the largest COMs for which the population is distributed over
many energy states, having large partition functions.
Figure 8 shows as an example the simulation of the 
HCOOCH$_3$ and HCOCH$_2$OH spectrum as expected for 
SKA1-MID modeled with GILDAS-Weeds
(Maret et al. 2011) in the range $\sim$ 1--15 GHz, assuming LTE
conditions and physical parameters expected in hot-corinos
(see e.g. J\"orgensen et al. 2012). 
In particular, if we consider bright lines (S$\mu^2$ $\geq$ 1 D$^2$) at low
excitation (E$_u$ $\le$ 20 K), then the 1--15 GHz band contains a considerable number of
COMs transitions. As an example the line densities for bright, low-excitation transitions of  
methyl formate and glycoaldehyde in the 1--15 GHz range is $\sim$ 1 line/GHz,
ten times higher than in the 80--300 GHz spectral window.

\begin{figure}
\centering
\includegraphics[width=1\textwidth]{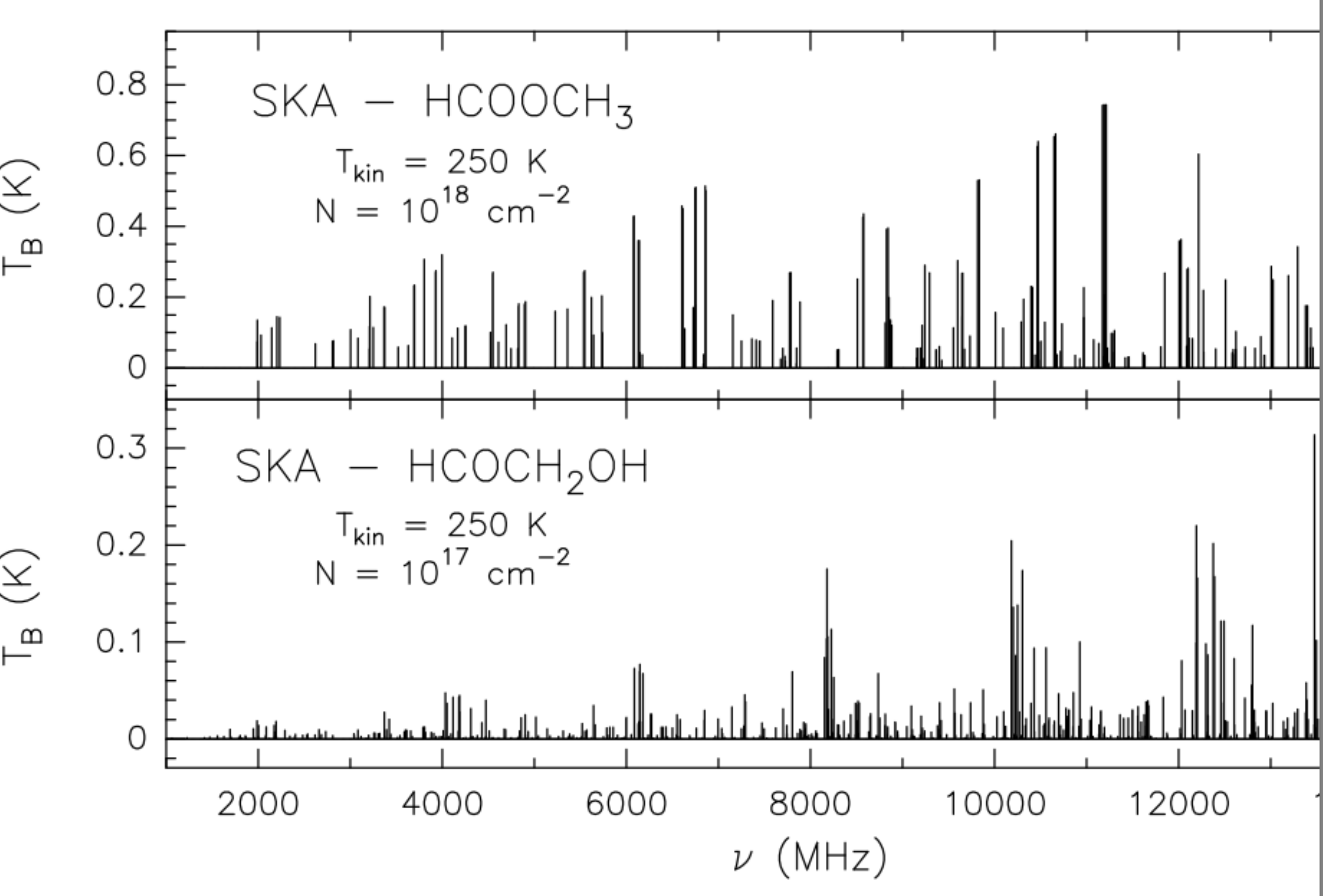}
\caption{Simulations of the spectrum of HCOOCH$_3$ and HCOCH$_2$OH as observed by
SKA1-MID modeled with GILDAS-Weeds
(Maret et al. 2011) in the range $\sim$ 1--15~GHz, assuming LTE
conditions and physical parameters expected for methyl formate and 
glycolaldehyde in hot-corinos
(see e.g. J\"orgensen et al. 2012): $T_{\rm kin} = 250$ K,
$N$(HCOOCH$_3$) = $10^{18}$ cm$^{-2}$, 
$N$(HCOCH$_2$OH) = $10^{17}$ cm$^{-2}$, source size = 1",
and line FWHM = 4 km s$^{-1}$.}
\label{glyco}
\end{figure}
 

An instructive view is given by the NRAO 100-m GBT PRIMOS Legacy Project,
which recorded the spectrum from 300 MHz to 46 GHz towards the Sgr B2(N)
molecular cloud. The PRIMOS data have resulted in numerous new detections
in astrochemistry (e.g. Loomis et al. 2013, and references therein).  The
COM emission is expected to come from regions at sub-arcsec scale (hot
corinos, jets, shocks).  SKA1 with band 5 operating at 5--14 GHz will reach
$\sim$35 mas resolution.  Therefore, the spatial resolutions offered by SKA
are fundamental to resolve and image the emitting region.  As a comparison,
the NRAO VLA interferometer, with 36 km maximum baseline, obtains a
synthetisized beam between 130 mas (15 GHz) and 2" (1 GHz)
(https://science.nrao.edu/facilities/vla/docs/manuals/oss).  In this way,
(i) we will have bright line emission, and (ii) we will correctly evaluate
COMs' abundances.  In addition, the sensitivity achieved by SKA for larger
beam sizes will be unbeatable, offering the possibility to detect the
emission from glycine and other amino acids in pre-stellar cores, where
their emission is expected to be relatively extended (around tens of
arcseconds).  To conclude, SKA offers a unique combination of spatial
resolution and high sensitivity to produce a complete inventory of known
interstellar species accessible in the centimeter wavelength range that can
be used to put severe constrains on the physical conditions of the emitting
gas, as well as on COMs abundances.

\vspace{0.5cm}

This work was partly supported by the Italian Ministero dell'Istruzione, 
Universit\`a e Ricerca (MIUR) through the grant {\it Progetti Premiali 2012 - iALMA}.

\bibliographystyle{apj}

\end{document}